\documentclass[a4paper,12pt]{article}

\usepackage{amssymb}
\usepackage{amsmath}
\usepackage[dvips]{graphicx}
\begin{document}

\begin{center}
{\bf  THE FEDOSOV MANIFOLDS AND MAGNETIC MONOPOLE}
\end{center}
\begin{center} {\it Alexander~S.~Ushakov} \end{center}
 St.Petersburg State University, V.A.Fock
Institute of Physics. Ulianovskaya ul. 1, Petrodvoretz,
Saint-Petersburg, Russia. asushakov@gmail.com

\begin{center} Abstract
\end{center}
{\it A non-symplectic generalization of Hamiltonian mechanics is
considered. It allows include into consideration "non-Lagrange"
systems, such as theory of charged particle in the field of
magnetic monopole. The corresponding generalization for the
Fedosov manifolds is given. The structure of phase space of
"charged particle in the field of magnetic monopole" is studied.}

\section{\bf Introduction}

There are equations of motion in mechanics, that cannot be
obtained in the framework of Lagrangian formalism, for example
~\cite[p.13]{Shabanov}:
\begin{equation}
\ddot q_i = \alpha e_{ijk} q_j \dot q_k, \ \ \ \ i,j,k=1,2,3,\quad
\alpha=const,
\end{equation}
where $e_{ijk}$ -- is absolutely antisymmetric tensor,
$e_{123}=1.$ The Hamiltonian formalism allows to deal with more
general class of dynamical systems. All the information about
system is contained in two objects -- the Hamilton function and
symplectic 2-form, while in Lagrangian mechanics only one function
(Lagrangian) is necessary. The action functional in the Lagrange
theory is varied on $n$ functions of time $q^i(t)$, but in
Hamiltonian formalism --- on $2n$ functions $(q^i(t),p_i(t))$.
Usually the simplest case is used in Hamiltonian mechanics, when
symplectic form can be transformed to standard one by the Darboux
transformation. These dynamical systems allows nontrivial
generalization by choosing different 2-form~\cite{Prokhorov}.
Equations (1) can obtained from the Hamiltonian equations of
generalized mechanics under the appropriate choice of 2-form .\\

To define a dynamical system, in Hamiltonian mechanics one needs:\\
 \hangindent=0.5cm \hangafter 2 \noindent
  1) Even-dimensional manifold $M^{2n}$ and the closed non-degenerated 2-form on it $\omega^2 = \sum_{\mu > \nu}\omega_{\mu
\nu}(x)\,\mbox{d}x^\mu\!\!\land \!\mbox{d}x^\nu,\,\,
\mu,\nu=1,2,...,2n,\,\, x\in M^{2n},$ $\mbox{d}\omega^2=~0.$ The
non-degeneracy of 2-form means that $\mbox{det}(\omega_{\mu
\nu}(x))\!\!\neq \!\!0$; and in this case the manifold $M^{2n}$ is orientable. The manifold $M^{2n}$ equipped with
 $\omega^2$ is called symplectic~\cite{Arnold, Fomenko}.\\
Differential 2-form allows us to define Poissons bracket for a
pair of functions $f(x),\,g(x):$
$$ \{f,g\}\equiv \frac{\partial f}{\partial
x^{\mu}}\, \omega ^{\mu \nu}  \frac{\partial g}{\partial x^{\nu}}
\equiv
 \omega ^{\mu \nu} \partial_{\mu}f \partial_{\nu}g\,,$$
 $$ \omega^{\mu \nu}\omega_{\nu \alpha}=\delta_{\alpha}^{\mu}.$$

 \hangindent=0.5cm \noindent
 2) Some function (Hamiltonian) of the local variables  $H(x)\equiv H(p,\,q) $ on $M^{2n}$, where $x =
(q^1, . . ., q^n, p_1, . . ., p_n)$, $q^i$ and $p_i$ -- are generalized coordinates and  momenta correspondingly.\\

\hangindent=0.5cm  \noindent
 3) One has to postulate the existence of "time" (positive parameter t);
  $q=q(t),\,p=p(t)$ are single--valued functions of $t$.\\

 \hangindent=0.5cm  \noindent
 4) One has to define equations of motion
 $$ \dot{x}^{\mu}= \{x^{\mu},H \}=\omega^{\mu \nu}\partial_{\nu}H\,,\,\,\,\,\dot{x}^{\mu}=\frac{\mbox{d}x^{\mu}}{\mbox{d}t}\,.$$

Besides obvious properties of Poissons brackets (linearity with
respect to each argument and antisymmetry) existence of the Jacoby
identity is supposed
\begin{equation}
\{f,\{g,h\}\} + \{g,\{h,f\}\} + \{h,\{f,g\}\}=0,
\end{equation}
that is equivalent to the fact that 2-form is closed

\begin{equation}
\partial_{\alpha}\omega_{\mu \nu} + \partial_{\mu}\omega_{\nu\alpha}
 + \partial_{\nu}\omega_{\alpha \mu } =0.
\end{equation}

 Notice, that the orientation of phase space defines "the arrow of time",
  namely, the Hamiltonian equations of motion are not invariant to reflection
  of time in contrast to the standard Lagrangian ones. Only
 simultaneous change of the matrix $\omega_{\mu\nu}$ (change the orientation) and the
 reversal of time does not change the equation of motion.

\par It is easy to obtain equation (1) by the simplest
generalization of Hamilton mechanics, namely by the proper choice
of 2-form. For a free particle ($H=\vec{p}^{\,2}/2$) one can
consider the antisymmetric $6\times6$ matrix
 \begin{equation}
\omega^{\,\mu\nu} = \left( \begin{array}{cc}
 ge_{ijk}p_k & \delta_{is} \\
 -\delta_{rj} & fe_{rsk}q_k
 \end{array} \right) \, ,
 \end{equation}
where $f=f(q),\,g=g(p)$ are arbitrary functions;
$i,j,r,s,k=1,2,3$. Then the Hamiltonian equations of motion
\begin{equation}
\dot{q}_i=p_i\,,\,\,\,\,\,\dot{p}_i=f e_{ijk}q_jp_k
\end{equation}
are equivalent to the equations (1), if $f=\alpha$. Equations (5)
with the specific choice of $f$ describe motion of charged
particle in the
field of magnetic monopole (see Sec. 2).\\

Deformation quantization \cite{Fedosov} is formulated in terms of
symplectic manifolds supplied with symmetric connection consistent
with the given symplectic structure. In the paper \cite{Gelfand},
devoted to study of the properties of symplectic curvature tensor,
the term "Fedosov manifolds" $ $ is proposed. We will use it both
in the case of non-closed 2-forms (non-symplectic manifolds) and
(as a result) in the case of
non-symmetric connections (see Sec. 3).\\

 In this paper we analyse generalization of the non-symplectic Hamiltonian mechanics.
 Structure of phase spaces is studied with the help of generalized Fedosov construction.
 Curvature tensor and torsion tensor of given phase space of "charged particle in the field of
 magnetic monopole" is not zero.\\

 In section 2 main peculiarities of non-symplectic Hamilton mechanics are formulated.
 It allows to obtain "non-Lagrange" $ $ equations of motion,
 particularly the motion of electric charged in the field of magnetic monopole.
In section 3 generalization of the Fedosov manifolds for
non-symplectic Hamilton mechanics is constructed. Section 4 is
dedicated to study of space phase structure, 2-form of which is
set by matrix (6). Corresponding 2-form allows to take into
consideration magnetic monopole.
Final remarks are given in the section 5.\\

\section{\bf Peculiarities of generalized Hamiltonian mechanics and magnetic monopole}

One of peculiarities of given generalization of Hamiltonian
mechanics is that 2-form defined by the matrix inverse to (4)
\begin{equation}
\omega_{\,\mu\nu} = \frac{1}{(1-fgq_kp_k)}\left( \begin{array}{cc}
 fe_{ijk}q_k & fgp_iq_s - \delta_{is} \\
 \delta_{rj}-fgq_rp_j & ge_{rsk}p_k
 \end{array} \right) \, ,
\end{equation}
is not closed: $\mbox{d}\omega^2\neq 0.$ So, one obtains
non-symplectic Hamiltonian mechanics. But what this peculiarity
leads to? Firstly, the Jacoby identity (2) is not satisfied, i.~e.
the Lie algebra  structure of the Poisson brackets on the surface
of smooth functions is broken. Secondly, the Darboux theorem is
inapplicable, since for its proof the Jacoby identity is needed
\cite{Fomenko}. Thirdly, the variational principle for these
theories is not known (for non-Lagrange systems see \cite{Gitman};
for inexact forms see paper \cite{Golovnev}, but closed form is
needed there (if one considers some star domain then any closed
form on it is exact); also see \cite{Kochan}). Fourthly, the
Hamiltonian phase flow does not preserve the Poincare invariants
besides phase volume (the Liouville theorem). To make sure of it
let us consider the element of phase space
$\Omega_k=(\omega^2)^k$, $k=1,2,...,n$; its differential
$$
\mbox{d}\Omega_k = k(\omega^2)^{k-1}\mbox{d}\omega^2
$$
is zero only in two cases: either form $\omega^2$ is closed or
$k=n$ ($\mbox{d}\Omega_n$ is $(2n+1)$-form, forms of this type are
absent in space
 $M^{2n}$, only if $\partial
\omega_{\mu\nu}/\partial t =0$; otherwise the Liouville theorem is
failed). Obviously, if $\mbox{d}\omega^2=0$, then $d\Omega_k /
dt=0$ irrespective of the choice of Hamiltonian,
what proves the theorem about the Poincare invariants  \cite{Arnold}.\\

Even if  $g=0$ in (6), one obtaines the same equations (5)
(Hamiltonian is the same), although 2-form has changed:

\begin{equation}
\omega^2=\mbox{d}p_i\!\!\land \!\mbox{d}q^i +
fe_{ijk}q^k\,\mbox{d}q^i\!\!\land \!\mbox{d}q^{\,j}.
\end{equation}

Equations (5) allow interesting interpretation. If $f =
eg_e/(mc(q^2_1+q^2_2+q^2_3)^{\frac{3}{2}}),\,q_i=r_i$, then one
obtains \cite{poincare}\footnote[1]{Poincare instead of $eg_e/mc$
used $\lambda$.}
\begin{equation}
\ddot{\vec{r}}=\frac{eg_e}{mc}\frac{\vec{r}\times\dot{\vec{r}}}{|r|^3}.
\end{equation}
Eq. (8) describes motion of the particle with charge $e$ and mass
$m$ in the field of magnetic monopole with magnetic charge $g_e$
($c$ -- is the velocity of light). But at the same time (8) are
the Hamiltonian equations of motion for free particle
($H=\vec{p}^{\,2}/2$) with modified symplectic form (7). Actually
in the case of $f = \lambda (q^2_1+q^2_2+q^2_3)^{\frac{3}{2}} $ ($
\lambda$ is some constant) 2-form (7) is closed everywhere, except
of one point: $d\omega^2=4\pi eg_e \delta (\vec{r})/(mc)$, where
$\delta(\vec{r})$ is Dirac delta function. Notice, that in paper
\cite{Nesterenko} magnetic monopoles are considered from the point
of view of string theory,
 and the variational principle is suggested.\\

 Equations (8) could be obtained not only by modification of symplectic structure.
 According to Cabibbo and Ferrari \cite{cabibbo}
 to include magnetic monopoles in the Maxwell theory one should present
 electromagnetic force tensor $F_{\mu\nu}$ in the form
\begin{equation}
F_{\mu\nu}=\partial_\mu A_\nu - \partial_\nu A_\mu +
e_{\mu\nu\rho\sigma}\partial_\rho B_\sigma,\atop
\mu,\nu,\rho,\sigma = 1,2,3,4,
\end{equation}
where $e_{\mu\nu\rho\sigma}$ -- is unit skew-symmetric tensor,
$e_{1234}=1$, and $B_\sigma$ -- is the second vector field.
Expression (9) comes to ordinary $F_{\mu\nu}=\partial_\mu A_\nu -
\partial_\nu A_\mu $ if
the last term vanishes ($B_\sigma = const$).\\

 The second vector potential $B_\mu$ increases the gauge transformation group. The variational principle
 is not found this case.\\

 In paper \cite{prokhush} equations (5) are obtained in terms of
 Hamiltonian mechanics with constrains and the standard symplectic matrix

$$
\omega^{\,\mu\nu} = \left( \begin{array}{cc}
 0 & \delta_{rs} \\
 -\delta_{rs} & 0
 \end{array} \right) \, , \quad r,s=1,2,3,4,5,6;
 $$
dimension of phase space here is $2n=12.$ Consider the Hamiltonian
$$
H= fe_{ijk}p_jq_kq_{i+3} + \frac{1}{2}\delta_{ij}\left( p_ip_j -
p_{i+3}p_{j+3} \right), \quad i,j,k=1,2,3,
$$
where $f=f(q_1,q_2,q_3)$. Equations of motion are:
$$
\dot{q}_i=fe_{ijk}q_jq_{k+3}+p_i,\quad
\dot{p}_i=fe_{ijk}p_jq_{k+3}+ \partial_{q_i}f\,e_{ijk}
q_ip_jq_{k+3},
$$
$$
 \dot{q}_{i+3}=-p_{i+3},\quad \dot{p}_{i+3}=fe_{ijk}p_kq_j,
\quad i,j,k=1,2,3.
$$
Let $q_i$ and $p_i$ obey the constraints
$$
\phi_i=p_i+p_{i+3}=0,\quad \phi_{i+3}=q_i-q_{i+3}=0, \quad
i=1,2,3.
$$
They are constrains of the 1st kind:
$$
\{\phi_i,H \}=fe_{ijk}q_jp_k + fe_{ijk}p_jq_{k+3}+
\partial_{q_i}f\,e_{ijk} q_ip_jq_{k+3}\approx0,
$$
$$
\{\phi_{i+3},H\}=fe_{ijk}q_jq_{k+3}+p_i+p_{i+3}\approx0, \quad
\{\phi_r,\phi_s\}=0,
$$
$$
i,j,k=1,2,3, \quad r,s=1,2,...,6.
$$
It is easy to see that equations of motion of physical variables
are indentical to Eq. (5). The Hamiltonian $H\approx0$ -- rather
 unusual property for mechanics with
constraints \cite[p.~140]{Shabanov}.\\

There exists a system similar to the previous one but with the
Hamiltonian
$$
\widetilde{H}=fe_{ijk}p_jq_kq_{i+3} + \frac{1}{2}\delta_{ij}\left(
p_ip_j + p_{i+3}p_{j+3} \right), \quad i,j,k=1,2,3,
$$
which is not equal to zero in the weak sence. Equations of motion
are the same except the equations $\dot{q}_{i+3}=p_{i+3}$. The
constrains read
$$
\widetilde{\phi}_i=p_i+p_{i+3}=0,\quad
\widetilde{\phi}_{i+3}=q^2_i-q^2_{i+3}=0, \quad i=1,2,3.
$$
They are constrains of 1st kind (no sum over i):

$$
\{\widetilde{\phi}_i,\widetilde{H} \}=fe_{ijk}q_jp_k +
fe_{ijk}p_jq_{k+3}+
\partial_{q_i}f\,e_{ijk} q_ip_jq_{k+3}\approx 0,
$$
$$
\{\widetilde{\phi}_{i+3},\widetilde{H}\}=2fq_ie_{ijk}q_jq_{k+3}+2q_ip_i-2q_{i+3}p_{i+3}\approx
0, \quad \{\widetilde{\phi}_r,\widetilde{\phi}_s\} \approx 0,
$$
$$
i,j,k=1,2,3, \quad r,s=1,2,...,6.
$$
Equations of motion for physical variables coincide with Eqs. (5).
Taking into consideration constraints we see that the Hamiltonian
$\widetilde{H}$ is not equal to zero. Notice that unlike theories
 considered usually each constraint $\widetilde{\phi}_{i+3} \approx 0$ have two different solutions.\\
\\

\section{\bf Fedosov's manifolds}

Let us consider a manifold $M^{2n}$ with non-degenerate 2-form
$\omega^2$. Let us determine covariant derivative $\nabla$ on
$M^{2n}$, consistent with 2-form $\omega^2$, i.e. $\nabla
\omega^2=0$ \cite{Fedosov}. In the coordinate basis this condition
means that
\begin{equation}
\nabla_k\omega_{\mu\nu}= \partial_k\omega_{\mu\nu} - \Gamma^l_{\mu
k}\omega_{l\nu}- \Gamma^l_{\nu k}\omega_{\mu l}=0.
\end{equation}

Let's introduce $\omega_{\nu l}\Gamma^l_{\mu k}=\Gamma_{\nu, \mu
k}$ and write Eq. (10) together with those, obtained by the cyclic
permutation of indices
\begin{equation}
 \partial_k\omega_{\mu\nu}= - \Gamma_{\nu, \mu k}+ \Gamma_{\mu, \nu k},
\end{equation}
\begin{equation}
 \partial_\mu\omega_{\nu k}= - \Gamma_{k, \nu\mu}+ \Gamma_{\nu, k\mu}.
\end{equation}
\begin{equation}
 \partial_\nu\omega_{k \mu}= - \Gamma_{\mu, k\nu}+ \Gamma_{k, \mu\nu},
\end{equation}

Adding expressions (12) и (13), and subtracting (11), one obtains
the following representation for $\Gamma_{k, \mu\nu}$:

\begin{equation}
\Gamma_{k, \mu\nu}= \frac{1}{2} \big( \partial_\nu\omega_{k\mu} +
\partial_\mu\omega_{\nu k} - \partial_k\omega_{\mu\nu}\big) + S_{\mu\nu k}+ S_{k\mu\nu} - S_{\nu
k\mu},
\end{equation}
where $S_{k\mu\nu}$ is symmetric part of connection $\Gamma_{k,
\mu\nu}$

\begin{equation}
S_{k\mu\nu}=\frac{1}{2}\big(\Gamma_{k, \mu\nu} + \Gamma_{k,
\nu\mu}\big).
\end{equation}

If 2-form $\omega^2$ is not closed (e.g., if it is given by matrix
(6)), then condition (10) imposes a restriction on symmetry of
indices of symplectic connection $\Gamma_{k, \mu\nu}$, namely,
connections cannot be symmetric. To make sure of this fact let's
us sum up equations (11)--(13) and take into consideration the
fact that the identity (3) is not satisfied

\begin{equation}
\partial_k\omega_{\mu\nu}+ \partial_\nu\omega_{k \mu}+ \partial_\mu\omega_{\nu
k}= 2T_{k\mu\nu} + 2T_{\mu\nu k} + 2T_{\nu k\mu}\neq 0,
\end{equation}
where $T_{k\mu\nu}=\frac{1}{2}\big(\Gamma_{k,\mu\nu} -
\Gamma_{k,\nu\mu}\big)$ -- is the antisymmetric part of connection
$\Gamma_{k, \mu\nu}$. From (16) it follows that connections could
not
be symmetric.\\

As it has been pointed in introduction we will use term the
Fedosov manifolds also to manifolds with given non-closed 2-form.\\

 Curvature tensor $R^i_{qkl}$ is determined by action of the commutator of covariant
  derivatives $[\nabla_k,\nabla_l ]= \nabla_k\nabla_l -
 \nabla_l\nabla_k$ on a vector field $a^i$. The first term reads:

 $$\nabla_k\nabla_l a^i=\nabla_k\big( \partial_la^i + \Gamma^i_{ql}a^q\big)=  $$
 $$= \partial_k\big(  \partial_la^i + \Gamma^i_{ql}a^q\big) + \Gamma^i_{pk}
 \big( \partial_la^p + \Gamma^p_{ql}a^q\big) - \Gamma^p_{lk}
 \big( \partial_pa^i + \Gamma^i_{qp}a^q\big) = $$
 $$\partial_k\partial_l a^i + \Gamma^i_{ql}\partial_k a^q + \Gamma^i_{pk}\partial_l a^p -
  \Gamma^p_{lk}\partial_p a^i + a^q \partial_k \Gamma^i_{ql} + \Gamma^i_{pk}\Gamma^p_{ql}a^q -
  \Gamma^p_{lk}\Gamma^i_{qp}a^q.$$

Then the commutator gives
\begin{equation}
[\nabla_k,\nabla_l ]a^i = -R^i_{qkl}a^q + T^p_{kl}\partial_pa^i,
\end{equation}
where

\begin{equation}
R^i_{qkl} = \partial_l \Gamma^i_{qk} - \partial_k \Gamma^i_{ql} +
\Gamma^i_{pl}\Gamma^p_{qk} - \Gamma^i_{pk}\Gamma^p_{ql},
\end{equation}
is the curvature tensor;

\begin{equation}
T^p_{kl}= \Gamma^p_{kl} - \Gamma^p_{lk}
\end{equation}
is the torsion tensor.\\

 Curvature tensor is antisymmetric with respect to lower indices
 $R^i_{qkl}=-R^i_{qlk}$.

 Let us introduce tensor $R_{jqkl}=\omega_{ji}R^i_{qkl}$, which is also antisymmetric
 with respect to lower indices
 \begin{equation}
R_{jqkl}=-R_{jqlk}.
 \end{equation}
  From the definition of $\Gamma_{j,qk}$ and expression (11) it follows that

 $$\omega_{ji}\partial_l\Gamma^i_{qk}=\partial_l\Gamma_{j,qk}-\partial_l\omega_{ji}\Gamma^i_{qk}=
 \partial_l\Gamma_{j,qk} + \big(\Gamma_{i,jl}- \Gamma_{j,il}\big)\Gamma^i_{qk}.$$

 So, taking into consideration (18) one gets

 \begin{equation}
R_{jqkl}= \partial_l\Gamma_{j,qk} -\partial_k\Gamma_{j,ql} +
\Gamma_{i,jl}\Gamma^i_{qk} - \Gamma_{i,jk}\Gamma^i_{ql}.
 \end{equation}

Using (11) one obtains following relations

$$\partial_l\partial_k\omega_{jq}=\partial_l\Gamma_{j,qk} - \partial_l\Gamma_{q,jk}, $$
$$ \partial_k\partial_l\omega_{jq}=\partial_k\Gamma_{j,ql} - \partial_k\Gamma_{q,jl}.$$

The latter equations lead to
\begin{equation}
\partial_l\Gamma_{j,qk} - \partial_k\Gamma_{j,ql} = \partial_l\Gamma_{q,jk} -
\partial_k\Gamma_{q,jl}.
\end{equation}

Using relation (22) and equality

$$\Gamma_{i,jl}\Gamma^i_{qk}=-\omega_{si}\Gamma^s_{jl}\Gamma^i_{qk}=-\Gamma^s_{jl}\Gamma_{s,qk}, $$
one obtains symmetry property of tensor $R_{jqkl}$  in respect to
the first two indices:

\begin{equation}
R_{jqkl}=R_{qjkl}.
\end{equation}
From (23) it follows that

$$R^i_{ikl}=\omega^{ip}R_{pikl}=0. $$

 Let us now introduce the Ricci tensor for Fedosov's manifold:

\begin{equation}
R_{ql} = R^i_{qil}=\omega^{ik}R_{kqil}.
\end{equation}

Notice, that in the case of symmetric connections Ricci tensor is
also  symmetric \cite{Gelfand}.\\

Let us introduce scalar curvature for Fedosov manifold:

\begin{equation}
R =\omega^{ik}R_{ki}.
\end{equation}

{\bf Theorem 1.} For any Fedosov's manifold scalar curvature $R=0$.\\
 {\bf Proof.} Using the definitions (24) and (25)
 $$R=\omega^{lk}R_{kl}=\omega^{lk}\omega^{ij}R_{jkil}=\frac{1}{2}\big(\omega^{lk}\omega^{ij}R_{jkil}+
 \omega^{lk}\omega^{ji}R_{ikjl} \big), $$
 and taking into account symmetry proporties of the curvature tensor (20) и (23),
 one obtains
 $$R=\frac{1}{2}\big(\omega^{lk}\omega^{ij}R_{jkil}-
 \omega^{lk}\omega^{ji}R_{kilj} \big)\equiv 0. $$
 The theorem is proved.\\

\section{\bf Magnetic monopole}

Now we can begin study of the phase space structure of such systems
as charged particle in the field of magnetic monopole. Firstly let's
consider 2-form given by matrix (6). From Sec.3 we know that space
has torsion and zero scalar curvature. The curvature tensor is not
uniquely defined by the connection consistent with the 2-form. The
symmetric part of the connection doesn't depend on the choice of the
2-form, hence one can consider only the part of the curvature tensor
associated with the skew-symmetric part of the connection.

{\bf Theorem 2.} If 2-form $\omega^2$ is non-degenerate, then
there exists unique connection that is skew-symmetric and
consistent with this 2-form $\omega^2$. This connection in
coordinate bases is given by:
\begin{equation}
\Gamma^l_{\mu\nu}= \frac{1}{2}\omega^{lk} \big(
\partial_\nu\omega_{k\mu} +
\partial_\mu\omega_{\nu k} - \partial_k\omega_{\mu\nu}\big).
\end{equation}

{\bf Proof.} Rising index $k$ in expression (14) (that is
consequence of consistency of connection with 2-form (10)) and
taking into consideration that
$\Gamma_{k,\mu\nu}=-\Gamma_{k,\nu\mu}$, one obtains required
formula. Theorem is proved.\\

Notice that the symmetric part of the connection doesn't transform
as tensor under the coordinate transformations, hence in new
coordinates the connection, in general, can have the symmetric part.

Consider the specific coordinate bases in which the connection is
skew-symmetric. Then one can calculate the curvature in this bases
by (21). It is the tensor with respect to the coordinate
transformations, hence the curvature in some other coordinate bases
can be obtained from the curvature in the chosen specific bases by
the tensor law, irrespective of possible break of the
skew-symmetricity of the connection.

If we consider the model with skew-symmetric connections only, some
additional constraints should be imposed. For example, we can chose
some specific coordinate system and allow only linear
transformations, or equip the manifold with the Riemann
(Pseudo-Riemann) metrics in addition to the 2-form. This metric can
be coordinated with the 2-form by the almost complex structure which
always exists on the even-dimensional orientable manifold. Although
metrics and almost complex structure is connected with the 2-form in
non-invariant way, it is convenient to use them together
\cite{Arnold}.  Then we require the connection to be coordinated
both with the 2-form and metrics. Then the metrics will be connected
with the symmetric part of the connection, while the 2-form - with
the skew-symmetric part.
%
%

 Let's show that in this case curvature tensor is not zero. It is
sufficient to prove this fact for any element of $R_{ijkl}$. We
will consider $R_{1112}$. Taking into account (26) and (21) one
has
\begin{multline}
R_{1112}=-\partial_{q_1}\Gamma_{112}=-\partial_{q_1}\partial_{q_1}\omega_{21}=
-\partial_{q_1}\left(\frac{q_3(\partial_{q_1}f +
f^2gp_1)}{(1-fgq^ip_i)^2} \right)=\\
=q_3\cdot\frac{\partial_{q_1}\partial_{q_1}f\cdot (1-fgq^ip_i)+
2(\partial_{q_1}f)^2g(q^ip_i)+3\partial_{q_1}f\cdot
fgp_1+f^3g^2p_1^2}{(1-fgq^ip_i)^3},
\end{multline}
where in the case of magnetic monopole
\begin{equation}
f=(q_1^2+q_2^2+q_3^2)^{-\frac{3}{2}}, \quad
g=(p_1^2+p_2^2+p_3^2)^{-\frac{3}{2}},
\end{equation}
$$ \partial_{q_1}f=
\frac{-3q_1}{(q_1^2+q_2^2+q_3^2)^{\frac{5}{2}}}, \quad
\partial_{q_1}\partial_{q_1}f=\frac{3(4q_1^2-q_2^2-q_3^2)}{(q_1^2+q_2^2+q_3^2)^{\frac{7}{2}}}.
$$

 For the Ricci tensor in case of 2-form (6), taking into consideration (4) and (24)
 one obtains
\begin{multline}
R_{kl}=\omega^{ij}R_{jkil}=gp_3(R_{2k1l}-R_{1k2l})+gp_2(R_{1k3l}-R_{3k1l})+gp_1(R_{3k2l}-R_{2k3l})+\\
+(R_{4k1l}-R_{1k4l})+(R_{5k2l}-R_{2k5l})+(R_{6k3l}-R_{3k6l})+\\
+fq_3(R_{5k4l}-R_{4k5l})+fq_2(R_{4k6l}-R_{6k4l})+fq_1(R_{6k5l}-R_{5k6l}).
\end{multline}

We are not going to determine all elements of Ricci tensor (29),
and limit ourselves only by diagonal elements $R_{ll}$. Using (21)
and definition $\Gamma_{r,kl}=\omega_{ri}\Gamma^i_{kl}$ we have:
\begin{equation}
R_{ll}=\omega^{kj}\partial_l \Gamma_{j,lk} +
\omega^{kj}\omega^{ir}\Gamma_{i,jl}\Gamma_{r,lk}.
\end{equation}
According to (26) the first term here is
\begin{multline}
\omega^{kj}\partial_l \Gamma_{j,lk} =
\frac{1}{2}\omega^{kj}\partial_l \left( \partial_k \omega_{jl} +
\partial_l \omega_{kj} - \partial_j \omega_{lk} \right)=\\
\frac{1}{2}\omega^{kj}\partial_l \partial_l \omega_{kj} +
\frac{1}{2}\left(\omega^{kj}\partial_l \partial_k \omega_{jl} -
\omega^{jk} \partial_l \partial_j \omega_{kl}\right) =
\frac{1}{2}\omega^{kj}\partial_l \partial_l \omega_{kj}.
\end{multline}

From $\omega^{ri}\omega_{ij}=\delta^r_j$ it follows
\begin{equation}
\omega^{ri}\partial_l \omega_{ij}=-\omega_{ij}\partial_l
\omega^{ri}.
\end{equation}
To find the second term (30) one uses Eqs. (26) and (32):
$$
\omega^{kj}\omega^{ir}\Gamma_{i,jl}\Gamma_{r,lk}=\frac{1}{4}\omega^{kj}\left(-\omega_{ij}\partial_l
\omega^{ri} + \omega^{ri}\partial_j \omega_{li}+
\omega_{jl}\partial_i \omega^{ri}\right)\Gamma_{r,kl}=
$$
$$
=\frac{1}{4}\left( \delta^k_i \partial_l \omega^{ri} +
\omega^{kj}\omega^{ri}\partial_j \omega_{li} + \delta^k_l
\partial_i\omega^{ri}\right)\left( \partial_l
\omega_{rk} + \partial_k \omega_{lr} - \partial_r
\omega_{kl}\right)=
$$
$$
=\frac{1}{4}\left( \partial_l \omega^{rk}\partial_l \omega_{rk}+
\omega^{kj}\omega^{ri}\partial_j \omega_{li}\left( \partial_l
\omega_{rk} + \partial_k \omega_{lr} - \partial_r
\omega_{kl}\right) \right)=
$$
$$
=\frac{1}{4}\left( \partial_l \omega^{rk}\partial_l \omega_{rk}+
\omega^{kj}\omega_{rk}\partial_l \omega^{ir}\partial_j \omega_{li}
-\partial_r \omega^{jk}\partial_j
\omega^{ri}\omega_{li}\omega_{lk}\right)=
$$
\begin{equation}
=\frac{1}{4}\left((\omega_{rk}\partial_l \omega^{rk})^2 +
\partial_l \omega^{ir}\partial_r \omega_{li}-\partial_r \omega^{jk}\partial_j
\omega^{ri}\omega_{li}\omega_{lk}\right).
\end{equation}

Substituting (31) and (33) into (30) we finally obtain

\begin{equation}
R_{ll}=\frac{1}{4}\left(2\omega^{rk}\partial_l
\partial_l \omega_{rk}+(\omega_{rk}\partial_l \omega^{rk})^2 +
\partial_l \omega^{ir}\partial_r \omega_{li}-\partial_r \omega^{jk}\partial_j
\omega^{ri}\omega_{li}\omega_{lk}\right).
\end{equation}
Notice that the expression (34) holds for any Fedosov's manifold
with skew-symmetric connection.\\

Then one can find explicit expression for the first diagonal
element of the Ricci tensor $R_{11}$ for 2-form given by (6). We
omit simple but tedious calculations; the final result is

\begin{multline}
2\omega^{rk}\partial_1
\partial_1 \omega_{rk} = -4g\cdot\frac{\partial_{q_1}\partial_{q_1}f\cdot
(q^ip_i)}{1-fgq^ip_i}-\\
-8g\cdot\frac{\partial_{q_1}f\big(\partial_{q_1}f\cdot g(q^ip_i)^2
+ fg(q^ip_i)p_1+p_1\big) + f^2gp^2_1}{(1-fg(q^ip_i))^2}\,\,;
\end{multline}
\begin{equation}
(\omega_{rk}\partial_1
\omega^{rk})^2=4g^2\cdot\frac{\big(\partial_{q_1}f\cdot(q^ip_i)+fp_1\big)^2}{(1-fg(q^ip_i))^2}\,\,;
\end{equation}
\begin{equation}
 \partial_1 \omega^{ir}\partial_r \omega_{1i}=\frac{f\big(\partial_{q_1}f\cdot(q^ip_i)
 -fp_1\big)\big( q_3\partial_{p_2}g-q_2\partial_{p_3}g\big)}{(1-fg(q^ip_i))^2}\,\,;
\end{equation}
\begin{equation}
\partial_r \omega^{jk}\partial_j
\omega^{ri}\omega_{1i}\omega_{1k}=\frac{2f\big(\partial_{q_1}f+f^2gp_1
\big)\big( p_2q_2+p_3q_3\big)\big(
q_3\partial_{p_2}g-q_2\partial_{p_3}g\big)}{(1-fg(q^ip_i))^2}.
\end{equation}
Substituting expressions (35)--(38) into (34) one finally obtains
\begin{multline}
R_{11}=-g\cdot\frac{\partial_{q_1}\partial_{q_1}f\cdot
(q^ip_i)}{1-fgq^ip_i}-g\cdot\frac{\partial_{q_1}f\big(\partial_{q_1}f\cdot
g(q^ip_i)^2 +2p_1\big) + f^2gp^2_1}{(1-fg(q^ip_i))^2}+\\
+f\cdot\frac{\Big( q_2\partial_{p_3}g-q_3\partial_{p_2}g\Big)
\Big(\partial_{q_1}f\big(p_1q_1-p_2q_2-p_3q_3\big) -fp_1\big(
1+2fg(p_2q_2+p_3q_3)\big) \Big)}{4(1-fg(q^ip_i))^2}\,,
\end{multline}
 where in the case of magnetic monopole $f$ and $g$ are given by Eqs. (28). To make sure that $R_{11}$
 is not zero lets consider the case
$p_1=q_1=p_3=q_3=0$, then (39) reads
\begin{equation}
R_{11}=\frac{3}{q_2^2(1-q_2^2p_2^2)}.
\end{equation}

 As it was mentioned before, if $g=0$ in (6), then (Hamiltonian is the same) one comes
 to the same equations (5),  although 2-form now is given by Eq. (7). In this case the structure
 of phase space is more simple.  There are only eighteen non-zero
 connections (we consider the case of magnetic monopole):
 \begin{equation}
\Gamma_{i,jk}=\frac{\varepsilon_{ijk}(2q_i^2-q_j^2-q_k^2) +
\delta_{ij}\varepsilon_{rik}q_rq_i}{(q_1^2+q_2^2+q_3^2)^{5/2}}\,,
 \end{equation}
where $i,j,k=1,2,3$; $\Gamma_{i,jk}=-\Gamma^{i+3}_{jk}$.\\

 Curvature tensor has 54 non-zero elements:
\begin{multline}
R^{i+3}_{skl}=\Big(
15\big[\delta_{il}\delta_{sk}\big]_{i\rightarrow
s}\varepsilon_{lkr}q_rq_lq_k
+3\delta_{is}\varepsilon_{skl}q_s (-2q^2_s + 3q^2_k+3q^2_l)+\\
+3\varepsilon_{rlk}q_r\big[\delta_{is}\delta_{sk}(4q_s^2-
q^2_r-q^2_l)\big]_{l \rightarrow k}+\\
 +3\big[\varepsilon_{ilk}(\delta_{sk}+
\delta_{sl})q_s(4q_i^2 -q_l^2- q_k^2)\big]_{i \rightarrow s}
\Big)\big(q_1^2+q_2^2+q_3^2 \big)^{-7/2},
 \end{multline}
where $i,s,k,l = 1,2,3;$ and $\big[a_{ijl}\big]_{i\rightarrow j}=a_{ijl}+a_{jil}$.\\

 All elements of the Ricci tensor are zero.\\

\section{\bf Conclusion}

 The Lagrangian mechanics is included in the Hamiltonian one as the special case (with the help of the Legandre transformation);
  generally the Hamiltonian mechanics radically differs from
the Lagrange one. There are twice as much variables in the
Hamiltonian mechanics as in the Lagrangian one; Hamiltonian and
independent 2-form
are defined in Hamiltonian mechanics.\\
It is impossible to obtain equation (1) in the framework of the
Lagrangian mechanics ~\cite{Shabanov}, although in the proper
choice of
 $\alpha$ if describes motion of charged particle in the field of magnetic monopole (8).
  In the simplest generalization of Hamiltonian mechanics
(by choosing 2-form) one can obtain equations (5), which are
equivalent to ~(1). This generalization is non-symplectic
($\mbox{d}\omega^2\neq 0$), i.e. the Jacoby identity is not
fulfilled for the Poissons brackets. One can present corresponding
2-form as the sum of standard symplectic form and some form
defined in the configuration space (7), that is equivalent to
inclusion of the field of magnetic monopole. There is no  Darboux
transformation in this theory, variational principle is not known,
but the theory is of interest because it allows us to
include into consideration "non-Lagrange" systems.\\

 The same equations (5) are obtained in the framework of standard Hamiltonian mechanics,
 owing to modification of phase space
$M^6\rightarrow M^{12}$ with introduction of non-physical
variables into the theory, i.~e. going theories with constrains.
At the same time there are two models: in the first one
Hamiltonian is zero
when taking constrains into consideration, in the second one it is non-zero.\\

Generalization of the Fedosov manifold into the case of non-closed
2-forms (non-symplectic manifolds) is constructed (see Sec.3). In
this case connections cannot be symmetric. In this case phase
space has non-zero torsion tensor (19). As in the
case of the standard Fedosov manifolds scalar curvature is zero.\\

Structure of phase space of such system as charged particle in the
field of magnetic monopole is studied (see Sec.4). Curvature
tensor for the given system is not zero (see Eq. (27)). The
structure of phase space depends on the choice between 2-form (7)
and 2-form given by matrix (6), although equations of motion (5)
do not change. In the first case all elements of the Ricci tensor
are zero, in the second one they are nonzero, see (39), (40).
Moreover, for 2-form (7) we find all nonzero connections (41) and
all nonzero
elements of the curvature tensor (42).\\

\paragraph{\bf Acknowledgment}

I am grateful to L.V. Prokhorov for useful comments and
suggestions.\\

\end{document}